\documentclass[prd,twocolumn,preprintnumbers,nofootinbib]{revtex4-2}

\usepackage{slashed}
\usepackage{amsmath,amssymb}
\usepackage{graphicx}
\usepackage{units}
\usepackage{bbold}
\usepackage{xcolor}
\usepackage{dsfont}
\usepackage[hyperfootnotes=false,colorlinks,citecolor=blue]{hyperref}

\usepackage{pifont}

\usepackage{comment}
\usepackage[normalem]{ulem}     

\usepackage{float}

\usepackage{orcidlink}
\newcommand{\beq}{\begin{equation}}
\newcommand{\eeq}{\end{equation}}
\newcommand{\bea}{\begin{eqnarray}}
\newcommand{\ena}{\end{eqnarray}}

\def \epsilon {\varepsilon}

\begin{document}

\title{Hints of a new leptophilic Higgs sector?}
\author{Yoav Afik$^{1,2,\orcidlink{0000-0001-8102-356X}}$, P. S. Bhupal Dev$^{3,\orcidlink{0000-0003-4655-2866}}$, Anil Thapa$^{4,\orcidlink{0000-0003-4471-2336}}$}
\affiliation{$^1$Enrico Fermi Institute, University of Chicago, Chicago, Illinois 60637, USA\\
$^2$Experimental Physics Department, CERN, 1211 Geneva, Switzerland\\
$^3$Department of Physics and McDonnell Center for the Space Sciences, Washington University, St. Louis, Missouri 63130, USA\\
$^4$Department of Physics, University of Virginia,
Charlottesville, Virginia 22904-4714, USA}

\hypersetup{
pdftitle={},
pdfauthor={ Anil Thapa}}

\begin{abstract}
We show that a new leptophilic Higgs sector can resolve some  intriguing anomalies in current experimental data across multiple energy ranges. Motivated by the recent CMS excess in the resonant $e\mu$ channel at 146~GeV, we consider a leptophilic two-Higgs-doublet model, and propose a novel resonant production mechanism for the neutral components of the second Higgs doublet at the LHC using the lepton content of the proton. Interestingly, the same Yukawa coupling $Y_{e\mu}\sim 0.65-0.81$ that explains the CMS excess also addresses the muon $(g-2)$ anomaly. Moreover, the new Higgs doublet also resolves the recent CDF $W$-boson mass anomaly. The relevant model parameter space will be completely probed by future LHC data.   
\end{abstract}

\maketitle

\section{Introduction}
Using the Higgs boson as the keystone for new physics searches is well-motivated~\cite{Dawson:2022zbb}, as an extended Higgs sector could potentially address some of the pressing issues plaguing the Standard Model (SM), including the gauge hierarchy problem, stability of the electroweak vacuum, mechanism of electroweak symmetry breaking, origin of the fermion masses and mixing, matter-antimatter asymmetry, and the nature of dark matter. Therefore, even though the measured properties of the 125-GeV Higgs boson discovered at the LHC~\cite{ATLAS:2012yve, CMS:2012qbp} are thus far consistent with the SM expectations~\cite{CMS:2022dwd, ATLAS:2022vkf}, further precision Higgs studies, as well as direct searches for additional Higgs bosons, must continue. 

An interesting aspect of beyond-the-SM (BSM) physics is lepton flavor violation (LFV), which is forbidden in the SM by an accidental global symmetry. In fact, the observation of neutrino oscillations~\cite{Super-Kamiokande:1998kpq,SNO:2001kpb,DoubleChooz:2011ymz,DayaBay:2012fng,RENO:2012mkc} necessarily implies LFV. However, despite intense experimental efforts, no corresponding LFV in the charged lepton sector has been observed~\cite{Calibbi:2017uvl}. Therefore, alternative searches for LFV involving exotic Higgs decays ($h\to e\mu, e\tau, \mu\tau$) could be powerful probes of BSM physics~\cite{Diaz-Cruz:1999sns, Blankenburg:2012ex, Harnik:2012pb, Crivellin:2015hha, Herrero-Garcia:2016uab,Endo:2020mev, Barman:2022iwj}. Both ATLAS and CMS Collaborations have performed such LFV Higgs searches with the $\sqrt s=13$~TeV LHC Run-2 data~\cite{ATLAS:2019old,CMS:2019pex, CMS:2021rsq, ATLAS:2023mvd, CMS:2023pte}. Although no evidence for LFV decays of the 125~GeV Higgs boson was found, CMS has reported an intriguing $3.8\sigma$ local ($2.8\sigma$ global) excess in the resonant $e\mu$ search around 146~GeV, with a preferred cross section of $\sigma(pp\to H \to e\mu)=3.89^{+1.25}_{-1.13}$~fb~\cite{CMS:2023pte}. If confirmed, this would be a clear sign of BSM physics. In this letter, we take the CMS $e\mu$ excess at face value and provide the {\it simplest} possible interpretation in terms of leptophilic neutral scalars within a two-Higgs-doublet model (2HDM). In this context, we propose a {\it novel} resonant production channel for the leptophilic neutral (pseudo)scalars at the LHC using the lepton parton distribution function (PDF) of the proton~\cite{Bertone:2015lqa, Buonocore:2020nai, Buonocore:2021bsf,Dreiner:2021ext}; see Fig.~\ref{fig:feynman}.  We show that this scenario can explain the CMS excess with a Yukawa coupling $Y_{e\mu}\sim 0.55-0.81$, while being consistent with all existing constraints. 

Another interesting feature of our solution is its intimate connection to two other outstanding anomalies in current experimental data, namely, the $(g-2)_\mu$ anomaly~\cite{Muong-2:2006rrc, Muong-2:2021ojo, Muong-2:2023cdq} and the CDF $W$-mass anomaly~\cite{CDF:2022hxs}.
We emphasize that the prospects of probing a leptophilic light Higgs sector at the energy and intensity frontiers is a worthwhile study in its own right, irrespective of the future status of these anomalies.

\section{Model Setup}
\begin{figure}[!t]
    \centering
    \includegraphics[scale=0.5]{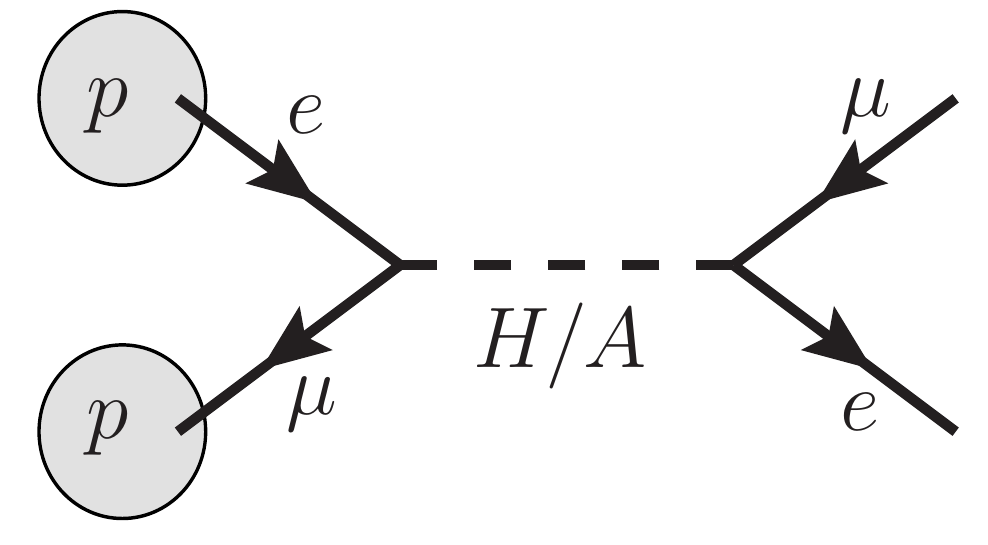}
    \caption{A representative Feynman diagram for resonant production of leptophilic scalar fields at hadron colliders through lepton PDF. }
    \label{fig:feynman}
\end{figure}
Here we propose an economical scenario with a leptophilic 2HDM to explain the CMS excess. We work in the Higgs basis~\cite{Davidson:2005cw}, where only one neutral Higgs acquires a nonzero vacuum expectation value, $v$. In this basis, the scalars fields can be parameterized as
\begin{equation}
H_{1}=\left(\begin{array}{c}
G^{+} \\
\frac{1}{\sqrt{2}}\left(v+H_{1}^{0}+i G^{0}\right)
\end{array}\right), \ H_{2}=\left(\begin{array}{c}
H^{+} \\
\frac{1}{\sqrt{2}}\left(H_{2}^{0}+i A\right)
\end{array}\right), \nonumber 
\end{equation}
where $(G^+,G^0)$ are the Goldstone modes, eaten up by $W$ and $Z$ after electroweak symmetry breaking, $(H^0_1,H^0_2)$ and $A$ are the neutral $\mathcal{CP}$-even and $\mathcal{CP}$-odd scalars respectively, and $H^+$ is a charged scalar field. In the alignment/decoupling limit~\cite{Gunion:2002zf, Carena:2013ooa, Dev:2014bir, Das:2015mwa}, we identify $H_1^0\equiv h$ as the observed $125$~GeV SM-like Higgs boson, whereas the $H_2$-sector does not couple to the SM gauge bosons. This is in agreement  with the LHC data~\cite{Haller:2018nnx, Eberhardt:2020dat, Karan:2023kyj}. We assume the mixing angle $\theta$ between the $\mathcal{CP}$-even scalar $H_2^0 \equiv H$ and the SM Higgs boson is small, and the only relevant production mechanism for $H$ (and $A$) at colliders is via its leptonic Yukawa interactions:
\begin{equation}
        - {\cal L}_Y \supset Y_{\alpha\beta} \bar{L}_\alpha H_2 \ell_{\beta,R} +  {\rm H.c.} 
        \label{eq:Yuk2HDM}
    \end{equation}
For either $Y_{e\mu}\neq 0$ or $Y_{\mu e}\neq 0$, with all other $Y_{\alpha\beta}$ involving electrons or muons assumed to be small, the dominant contribution to the $pp\to H/A\to e\mu$ signal comes from the $s$-channel Feynman diagram shown in Fig.~\ref{fig:feynman}, where the $H/A$ is produced resonantly using the lepton PDF of the proton, and then decays to $e^\mp\mu^\pm$  final states with a branching ratio (BR) determined by the structure of the Yukawa coupling matrix $Y$ in Eq.~\eqref{eq:Yuk2HDM}. There is a sub-dominant contribution to the same final-state from a $t$-channel exchange of $H/A$, not shown in Fig.~\ref{fig:feynman}, but included in our calculation. 

\begin{figure*}[!t]
    \centering
    \includegraphics[width=0.482\textwidth]{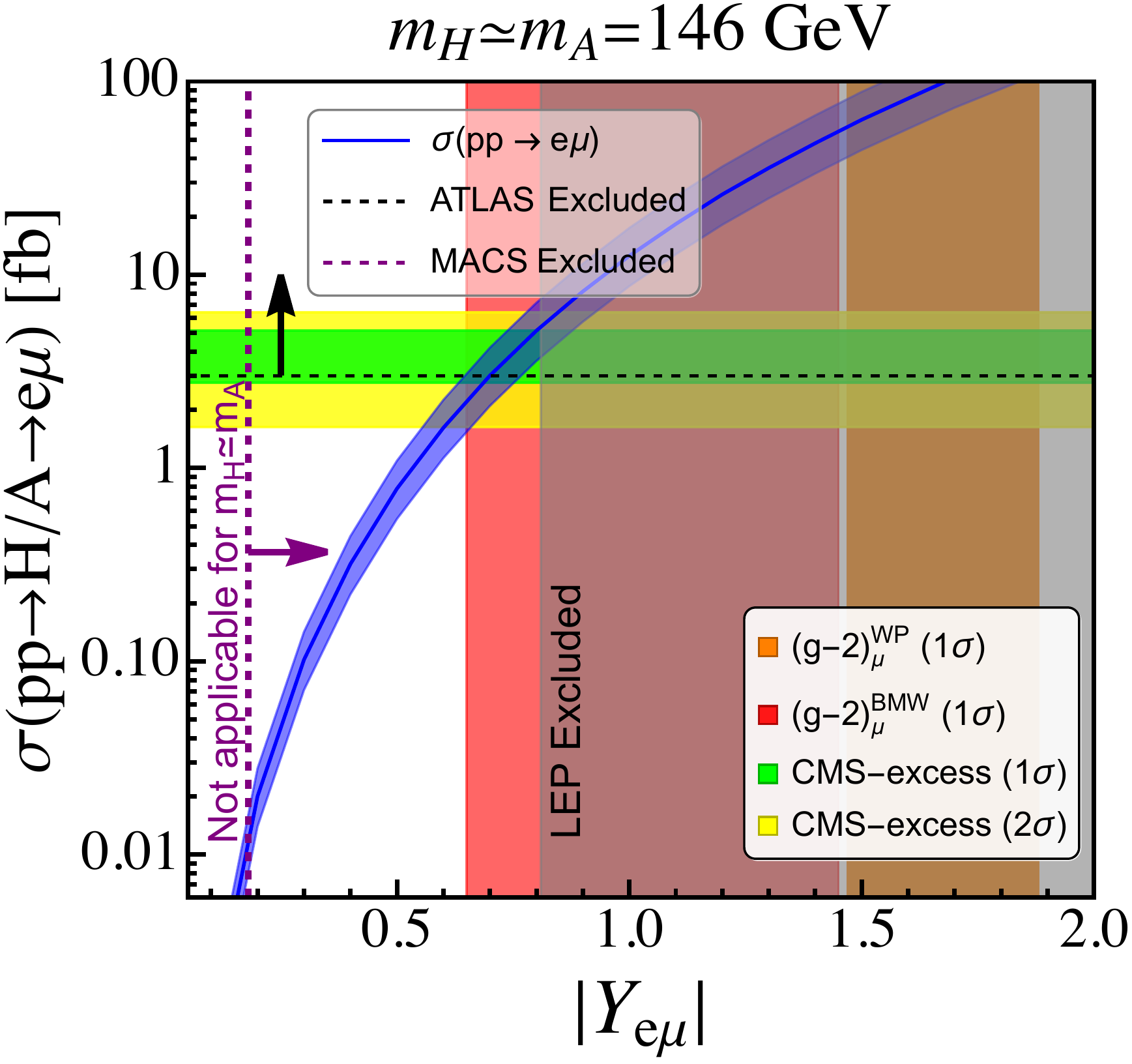} \hspace{2mm}
    \includegraphics[width=0.482\textwidth]{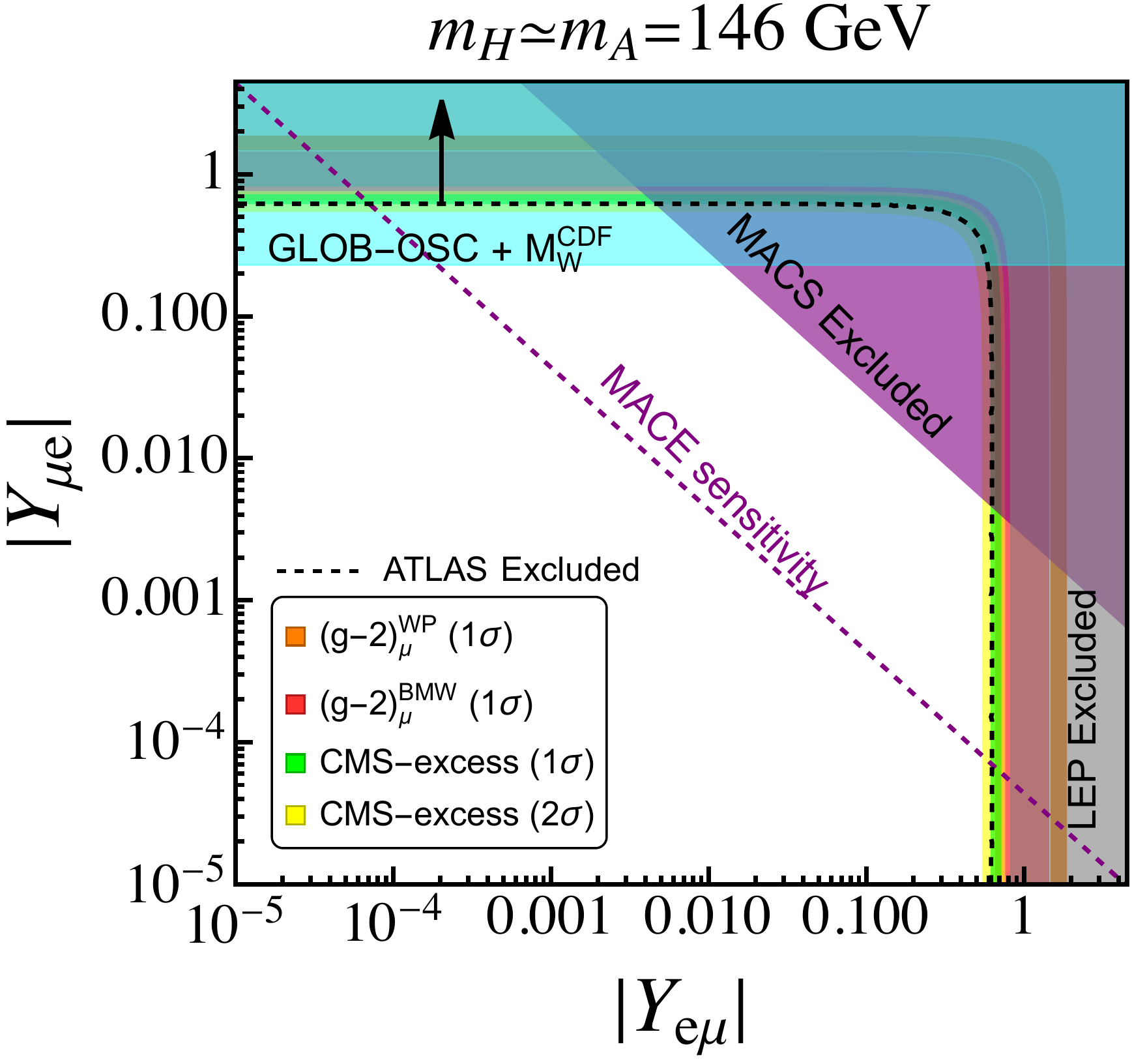}
    \caption{{\it Left:} Total $e\mu$ production cross section from $H/A$ (blue band) at $\sqrt s=13$~TeV LHC as a function of the Yukawa coupling $Y_{e\mu}$ (or $Y_{\mu e}$) in our leptophilic 2HDM with $m_H\simeq m_A=146$~GeV.  
    {\it Right:} Same as left panel but in the $Y_{e\mu}-Y_{\mu e}$ plane. See text for details. 
    } 
    \label{fig:p2}
\end{figure*}

We estimate the signal cross section numerically using {\sc MadGraph5\_aMC@NLO}~\cite{Alwall:2014hca} at leading order (LO) parton-level with the {\sc LUXlep-NNPDF31} PDF ({\sc 82400})~\cite{Manohar:2016nzj,Manohar:2017eqh,Bertone:2017bme,Buonocore:2020nai}. The default {\sc MadGraph5} cuts are applied at parton-level, and the default LO dynamical scale is used, which is the transverse mass calculated by a $k_t$-clustering of the final-state partons~\cite{Catani:1993hr}. The cross section result including both $H$ and $A$ contributions is shown by the blue curve in Fig.~\ref{fig:p2} left panel as a function of $|Y_{e\mu}|$ (also applicable for $|Y_{\mu e}|$) for $m_{H/A}=146$~GeV and assuming ${\rm BR}(H/A\to e\mu)=70\%$  (explained below), where the thickness accounts for the theory uncertainty due to scale ($^{+39.4\%}_{-30.3\%}$) and PDF ($\pm 4.5\%$) variation. The horizontal green (yellow) shaded region explains the CMS excess at $1\sigma$ ($2\sigma$). The corresponding ATLAS search~\cite{ATLAS:2019old} is not directly comparable with the CMS analysis, but a back-of-the-envelope calculation from the sideband data mildly disfavors a narrow-width excess at 146~GeV, and a rough scaling of background gives a ballpark upper limit of about 3.0~fb on the cross section~\cite{Leney}, as shown by the horizontal dashed line in Fig.~\ref{fig:p2}. We find that $Y_{e\mu}\sim 0.55-0.81$ can explain the CMS excess at $2\sigma$. For such values of the leptonic Yukawa coupling, any quark Yukawa couplings of the second Higgs doublet $H_2$ must be small; otherwise, it will be ruled out by the chirality enhanced meson decays, such as $\pi^+\to e^+\nu$. Thus our proposal is different from other scalar interpretations of the CMS excess~\cite{Primulando:2023ugc, Koivunen:2023led}, which used quark couplings to enhance the production cross section.

\section{Constraints}
The large $Y_{e\mu/\mu e}$ couplings of the neutral components, as well as the charged component, of the leptophilic Higgs doublet, are subject to a number of other constraints, and also give rise to other interesting phenomena, as discussed below. 

\subsection{Neutral sector}
Even if we choose only the off-diagonal entries $Y_{e\mu/\mu e}\neq 0$, small diagonal entries $Y_{\ell\ell}\sim \sin\theta y_\ell$ (with $\ell=e,\mu$) will be induced via the $h-H$ mixing and the SM Yukawa couplings $y_\ell\equiv \sqrt 2m_\ell/v$ (with $y_\mu\simeq 6\times 10^{-4}$ and $y_e\simeq 3\times 10^{-6}$). But the products $Y_{e\mu}Y_{ee}$ and $Y_{e\mu}Y_{\mu\mu}$ are subject to strong LFV constraints~\cite{Babu:2019mfe}. Using the general LFV formula~\cite{Lavoura:2003xp} and the current MEG limit on $\mu\to e\gamma$~\cite{MEG:2016leq}, we require $Y_{ee}\lesssim 9\times 10^{-5}$ and $Y_{\mu\mu}\lesssim 6\times 10^{-5}$, which gives an upper limit of $\sin\theta\lesssim 0.1$ on the Higgs mixing. 

The same $Y_{e\mu \ (\mu e)}$ coupling gives an additional contribution to the  $e^+ e^- \to \mu^+ \mu^-$ cross section via $t$-channel $H/A$ exchange, and therefore, is constrained by LEP measurements, which are in good agreement with the SM prediction~\cite{OPAL:2003kcu, Electroweak:2003ram}. Naively, the contact interaction bounds from LEP data would kill  the parameter space for ${\cal O}(1)$ Yukawa couplings~\cite{Babu:2019mfe}. However, this bound is not directly applicable, if neutral scalars are lighter than the LEP center-of-mass energy $\sqrt s=209$~GeV. A dedicated analysis~\cite{Barman:2021xeq} comparing the 2HDM cross section, which includes the interference between the $H/A$-mediated diagrams with the SM processes, against the LEP dimuon data imposes the constraint $Y_{e\mu}<0.8$, thus ruling out the parameter space shown by the brown-shaded region in Fig.~\ref{fig:p2}. The same bounds are also applicable to the $Y_{\mu e}$ coupling; see Fig.~\ref{fig:p1} for different masses. The LEP limit can be significantly improved at future lepton colliders, such as the $\sqrt{s}=1$~TeV ILC~\cite{Barklow:2015tja} with integrated luminosity $L=500~{\rm~fb}^{-1}$ (cf. the dashed curve in Fig.~\ref{fig:p1}), which can probe $Y_{e\mu}$ (or $Y_{\mu e}$) up to 0.1~\cite{Barman:2021xeq, Dev:2017ftk, Dev:2018vpr}. 

As for the hadron collider constraints on light neutral scalars, most of the Tevatron/LHC searches are done in the context of either MSSM or general 2HDM, and rely on the gluon fusion or vector boson fusion production mechanisms. None of these searches are applicable for us, because the leptophilic $H/A$ does not directly couple to the quarks, and in the alignment limit ($\theta\to 0$), also does not couple to the $W/Z$ bosons. This also suppresses other production channels like pair-production of $HA$. 

The most important constraint on the neutral scalar sector comes from low-energy process of muonium ($M_\mu=e^-\mu^+$)-antimuonium ($\overline{M}_\mu=e^+\mu^-$)  oscillation~\cite{Pontecorvo:1957cp,  Jentschura:1997tv, Clark:2003tv, Fukuyama:2021iyw}. The MACS experiment at PSI puts an upper bound on the oscillation probability $P(M_\mu\leftrightarrow\overline{M}_\mu) < 8.2 \times 10^{-11}$ at $90\%$ CL~\cite{Willmann:1998gd}, while a sensitivity at the level of $\mathcal{O}(10^{-14})$ is expected at the proposed MACE experiment~\cite{Bai:2022sxq}. In our 2HDM setup, the oscillation probability gets contribution from both $H$ and $A$~\cite{Conlin:2020veq, Fukuyama:2021iyw}; {see {\it Appendix A}. 
If $H$ and $A$ are highly non-degenerate, i.e. only either $H$ or $A$ dominantly contributes, the MACS bound requires $Y_{e\mu}<0.18$ for $m_{H/A}=146$~GeV, as shown (for illustration only) by the vertical purple line in Fig.~\ref{fig:p2} left panel, which rules out the LFV coupling needed to explain the CMS excess with a single scalar/pseudoscalar.  However, for $m_H\simeq m_A$,  there is a cancellation in the $M_\mu\leftrightarrow\overline{M}_\mu$ amplitude which allows for either $Y_{e\mu}$ or $Y_{\mu e}$ to be large, but not both. This is depicted by the gray-shaded region in Fig.~\ref{fig:p2} right panel for $m_H\simeq m_A=146$~GeV. In this limit, even the future MACE sensitivity cannot rule out the CMS excess region.

Thus far, it seems either $Y_{e\mu}$ or $Y_{\mu e}$ coupling can be taken to be large for explaining the CMS excess, while being consistent with the current constraints. However, as discussed below, a combination of the LHC charged Higgs constraints and the global fit to non-standard neutrino interactions (NSI), preclude the possibility of a large $Y_{\mu e}$ coupling, as shown by the horizontal purple-shaded region in Fig.~\ref{fig:p2} right panel. Therefore, the only viable possibility is to have a large $Y_{e\mu}$ coupling and small $Y_{\mu e}$ coupling (the lower right band of the CMS excess region in Fig.~\ref{fig:p2} right panel).    

\subsection{Charged sector}
    At LEP, $H^\pm$ can be pair produced through either $s$-channel Drell-Yan process via $\gamma/Z$, or $t$-channel via light neutrino. It can also be singly produced either in association with a $W$ boson or through the Drell-Yan channel in association with the leptons~\cite{Babu:2019mfe}. Once produced, the charged scalar decays into $\nu_\alpha \ell_{\beta,R}$ through the Yukawa coupling $Y_{\alpha \beta}$, which has the same signature as the right-handed slepton decay into lepton plus massless neutralino in SUSY models: $e^+e^- \to \tilde{\ell}_R^+ \tilde{\ell}_R^- \to \ell_R^+ \tilde{\chi}^0 \ell_R^- \tilde{\chi}^0$. We can therefore reinterpret the LEP slepton searches~\cite{ALEPH:2001oot,ALEPH:2003acj,DELPHI:2003uqw,L3:2003fyi,OPAL:2003nhx} to derive a bound on light charged scalars. Depending on the branching ratio ${\rm BR}(H^+\to \ell^+ \nu)$ the LEP limit on the charged scalar varies from $80-100$~GeV~\cite{Babu:2019mfe}.

Similarly at the LHC, a pair of charged scalars can be produced through $s$-channel Drell-Yan process via $\gamma/Z$, followed by decays into $\nu_\alpha \ell_{\beta,R}$. By reinterpreting the LHC searches for right-handed sleptons, one can therefore put bounds on the charged scalar mass as a function of BR in the massless neutralino limit. From an ATLAS analysis of the LHC Run-2 data~\cite{ATLAS:2019lff},
we obtain a lower bound of $m_{H^+}>425$~GeV at 90\% CL for ${\rm BR}(H^+\to \mu^+\nu_e)=1$. As we will see below, for $m_H=m_A=146$~GeV, the charged Higgs boson cannot be too much heavier due to the electroweak precision data (EWPD) constraints. Therefore, we would need additional decay channels in order to make ${\rm BR}(H^+\to \mu^+\nu_e)<1$ and relax the LHC constraints.

\section{Resolving the \texorpdfstring{$W$}{W}-boson mass anomaly}
The mass splitting between the neutral and charged components of the $SU(2)_L$ doublet $H_2$ breaks the custodial symmetry of the SM at the loop level. The change in the relationship between the $W$ and $Z$ boson masses can be used to accommodate the recent CDF $W$-mass anomaly, which currently stands at $7\sigma$~\cite{CDF:2022hxs}. This effect can be parameterized by the oblique parameters $S$ and $T$~\cite{Peskin:1990zt,Peskin:1991sw}, which modifies~\cite{Maksymyk:1993zm}
\begin{align}
        m_W \simeq m_W^{\rm SM} \left[1 - \frac{\alpha (S-2 \cos^2{\theta_w} T)}{4(\cos^2{\theta_w}-\sin^2{\theta_w})}  \right] , 
\label{eq:MW-STU}
\end{align}
where $\theta_w$ is the electroweak mixing angle. 
We incorporate the global electroweak fit~\cite{Lu:2022bgw} with the new CDF data to show allowed ranges for the scalar masses $(m_A, m_{H^+})$ with the choice of $m_H=146$~GeV in Fig.~\ref{fig:ST} (blue band). 
In spite of explaining the CDF $W$ mass shift, the model is mildly consistent with the PDG global fit~\cite{ParticleDataGroup:2022pth}, as can be seen from the red region in Fig.~\ref{fig:ST}.  We find that the CDF anomaly prefers significant splitting between $m_A$ and $m_{H^+}$. For $m_H=m_A=146$~GeV, we require $m_{H^+}\simeq$ 228--234 GeV to explain the CDF anomaly at $2\sigma$.  

\begin{figure}[t!]
    \centering
    \includegraphics[width=0.4\textwidth]{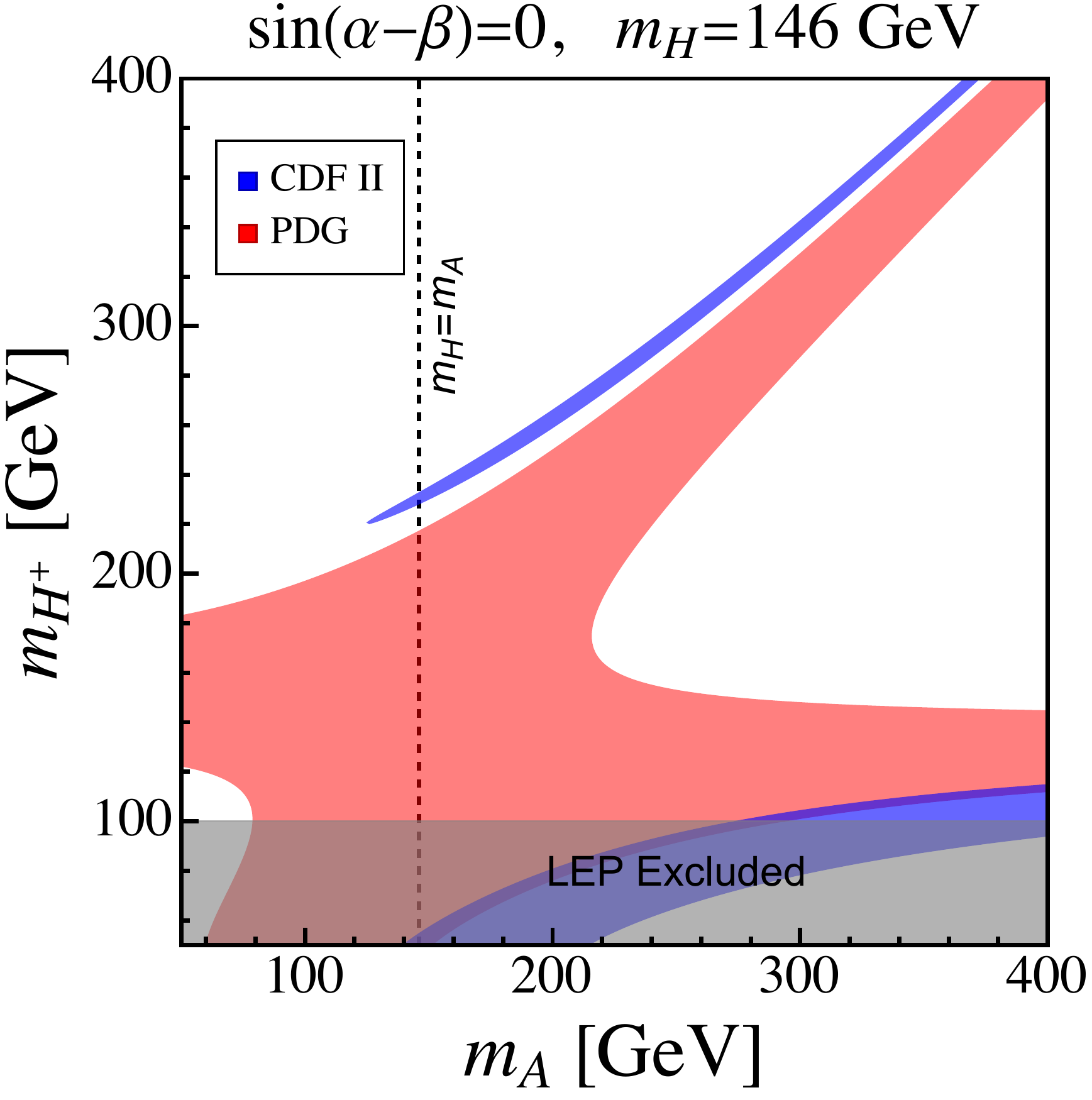}
    \caption{$2\sigma$ allowed ranges from EWPD global fit for the charged and neutral Higgs masses in the alignment limit of our 2HDM scenario. 
}
    \label{fig:ST}
\end{figure}

To reconcile the CDF-preferred $m_{H^+}$ region with the LHC constraint $m_{H^+}>425$~GeV, we reinterpret the slepton search limit as a function of the charged Higgs mass and ${\rm BR}(H^+\to \mu^+\nu_e)$, using the publicly available cross section limits given as a function of the slepton mass from the auxiliary material of Ref.~\cite{ATLAS:2019lff}, as well as from an earlier ATLAS analysis~\cite{ATLAS:2018ojr}. We find that to lower the $m_{H^+}$ bound to $\sim 230$~GeV, as required by the CDF anomaly, we need ${\rm BR}(H^+\to \mu^+\nu_e)<0.7$ (0.95) according to the cross section limits reported in Ref.~\cite{ATLAS:2018ojr} (\cite{ATLAS:2019lff}). We therefore fix  ${\rm BR}(H^+\to \mu^+\nu_e)=0.7$ for our analysis of the CMS excess in Fig.~\ref{fig:p2}. 

For the purpose of our discussion here, we are agnostic about the detailed structure of the Yukawa coupling matrix, which could account for the remaining 30\% BR. Additional nonzero entries in the Yukawa matrix are viable, albeit requiring potential adjustments to suppress LFV. One example texture that fits our branching ratio requirement is  $Y_{e\mu}=0.71$, $Y_{\tau\tau}=0.46$, and all other Yukawa entries negligible. This choice does not lead to trilepton LFV decays but does induce the radiative LFV decay $\mu \to e \gamma$ via a two-loop process involving the tauon in the Barr-Zee diagram~\cite{Ilisie:2015tra,Crivellin:2015hha}. However, it is also important to consider other diagrams such as the two-loop Barr-Zee diagram from the charged Higgs, which depends on the quartic coupling $\lambda (H_2^\dagger H_2) (H_1^\dagger H_2)$, and depending on the sign of $\lambda$, can destructively interfere with the tau-loop-induced diagram. We find that the LFV constraints can be satisfied for the above choice of Yukawa couplings for a relatively small quartic coupling of order $\mathcal{O}(10^{-3})$. 

We note here that instead of a large $Y_{e\mu}$ coupling, if we had allowed a large $Y_{\mu e}$ coupling, it would imply the coupling of charged Higgs $H^-$ to electrons and muon neutrinos. This leads to a $\nu_\mu-e$ coherent scattering in matter via $t$-channel exchange of the charged Higgs, and hence, generates an NSI of the type $\epsilon_{\mu\mu}=|Y_{\mu e}|^2/(4\sqrt 2 G_Fm_{H^+}^2)$~\cite{Babu:2019mfe}.  
From a recent global analysis of NSI constraints, we get a 90\% CL bound of $\epsilon_{\mu\mu}< 0.015$~\cite{Coloma:2023ixt}.\footnote{This is derived from the bound on $\epsilon_{\tau\tau}-\epsilon_{\mu\mu}$~\cite{Coloma:2023ixt} (see also Ref.~\cite{IceCube:2022pbe}), which is stronger than the individual bound on $\epsilon_{\mu\mu}$. In our model, both $\epsilon_{\mu\mu}$ and $\epsilon_{\tau\tau}$ cannot be simultaneously large due to strong charged LFV constraints; therefore, the bound on $\epsilon_{\tau\tau}-\epsilon_{\mu\mu}$ is also applicable for $\epsilon_{\mu\mu}$.} For $m_{H^+}\sim 230$~GeV, this gives an upper bound of $Y_{\mu e}\simeq 0.23$, which is shown by the purple-shaded region in Fig.~\ref{fig:p2} right panel.       

\section{Muon Anomalous Magnetic Moment}
The same $Y_{e\mu}$  coupling also contributes to the $(g-2)_\mu$ via the neutral and charged Higgs loops~\cite{Leveille:1977rc, Lindner:2016bgg}; see {\it Appendix B}. 
The combined result of the Brookhaven~\cite{Muong-2:2006rrc} and Fermilab~\cite{Muong-2:2021ojo} $(g-2)_\mu$ experiments is $4.2\sigma$ away from the 2020 global average of the SM prediction~\cite{Aoyama:2020ynm}: $\Delta a_\mu ({\rm WP}) = (251\pm 59)\times 10^{-11}$.\footnote{This was recently updated to $\Delta a_\mu ({\rm WP}) = (249\pm 48)\times 10^{-11}$~\cite{Muong-2:2023cdq}, but there is no noticeable change in our results.}
This discrepancy is however reduced to only $1.5\sigma$, if we use the ab-initio lattice calculation from the BMW collaboration~\cite{Borsanyi:2020mff}\footnote{Other lattice calculations now agree with the BMW result in the "intermediate distance regime"~\cite{Ce:2022kxy, ExtendedTwistedMass:2022jpw, Bazavov:2023has, Blum:2023qou}, but a more thorough and complete analysis is ongoing.}, which gives $ \Delta a_\mu ({\rm BMW}) = (107\pm 70)\times 10^{-11}$~\cite{Wittig:2023pcl}.   
The extra contribution from the neutral Higgs sector in our 2HDM scenario can explain the $(g-2)_\mu$ anomaly at $1\sigma$, as shown by the red (orange) shaded region in Fig.~\ref{fig:p2}, using the BMW (WP) value for the SM prediction. We find that the $1\sigma$ WP-preferred region is excluded by LEP constraint on $Y_{e\mu}$ for $m_H\simeq m_A=146$~GeV, whereas part of the $1\sigma$ BMW-preferred region is still allowed, while simultaneously explaining the CMS excess and the CDF $W$-mass anomaly.  

Fig.~\ref{fig:p1} shows the range of the $(g-2)_\mu$ anomaly-preferred region at $1\sigma$ in the neutral Higgs mass-coupling plane. For comparison, the green bar at 146~GeV shows the CMS excess region, whereas the purple shaded region around it is the exclusion region derived from CMS data~\cite{CMS:2023pte}. The gray-shaded region shows the LEP exclusion from $e^+e^-\to \mu^+\mu^-$ data~\cite{Barman:2021xeq}. The magenta region is excluded at $2\sigma$ from the precision $Z$-width measurements~\cite{ParticleDataGroup:2022pth}, because for $m_{H/A}<m_Z$, an additional decay mode $Z\to \ell_\alpha^+\ell_\beta^- H/A\to 4\ell$ opens up.  The vertical cyan (blue) line is the indirect lower bound on the neutral Higgs mass, derived using a combination of the electroweak precision constraint on the mass splitting between the neutral and charged Higgs sectors using the CDF (PDG) value of $m_W$, and the LEP lower limit of $\sim 100$~GeV on the charged Higgs mass. 
\begin{figure}[!t]
    \centering
    \includegraphics[width=0.48\textwidth]{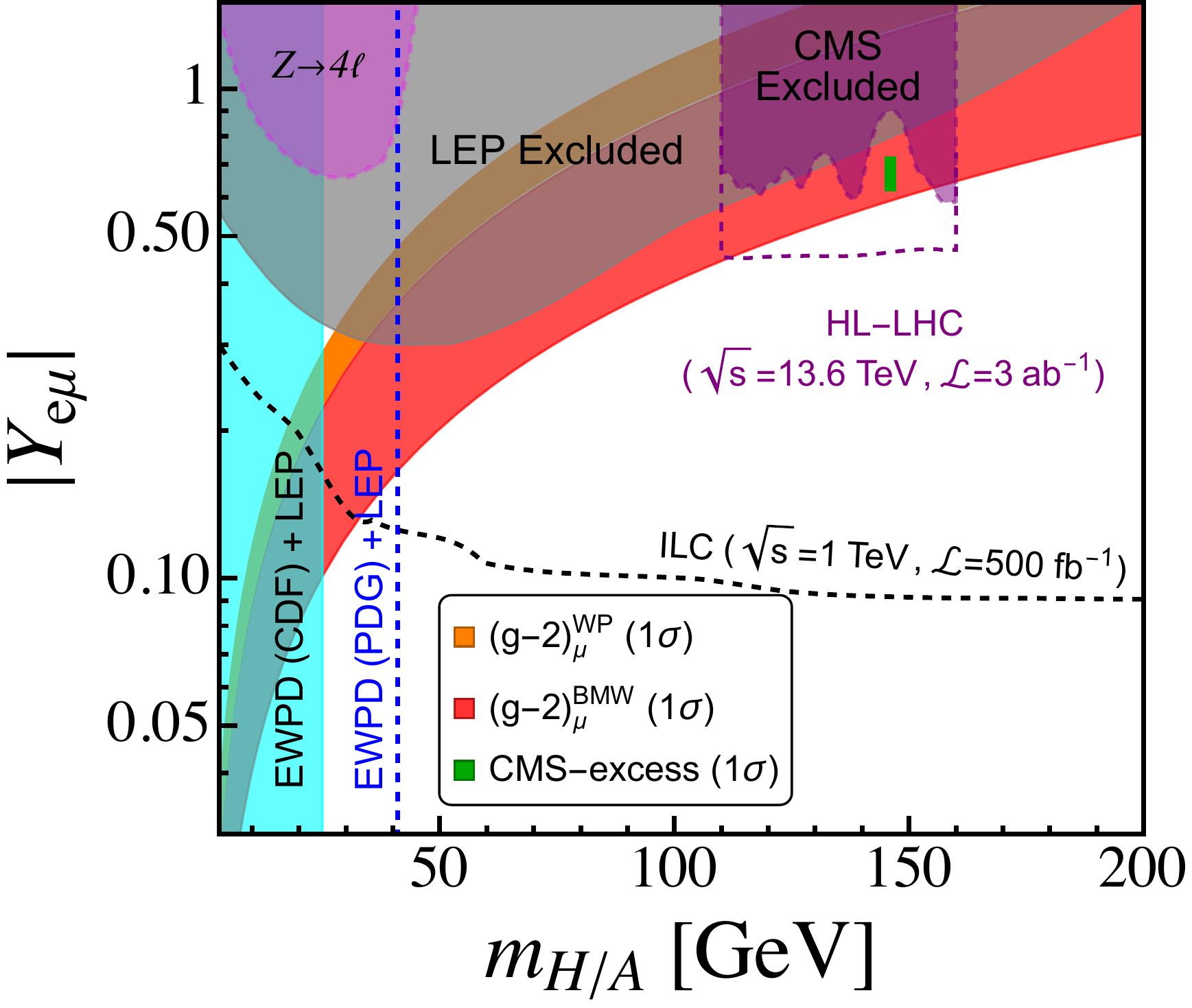}
    \caption{The CMS excess at $1\sigma$ (green) and 95\% CL exclusion (purple) in the mass-coupling plane, contrasted with the $1\sigma$ regions preferred by $(g-2)_\mu$. Also shown are the constraints from LEP dilepton, $Z\to 4\ell$, EWPD, and the future ILC and HL-LHC sensitivities.
    }
    \label{fig:p1}
\end{figure}
From Fig.~\ref{fig:p1}, we find that if we use the WP value for $g-2$, only a narrow band around $m_{H/A}\simeq 25$~GeV can explain the $g-2$ anomaly at $1\sigma$. On the other hand, if we use the BMW value, most of the parameter space for $m_{H/A}>25$~GeV is currently allowed. Future sensitivity projections from HL-LHC~\cite{ZurbanoFernandez:2020cco} and ILC~\cite{Barklow:2015tja} can cover most of the remaining allowed parameter space, irrespective of the status of the CMS excess. In general, a dedicated neutral scalar search in the LFV dilepton channels beyond 160~GeV could completely probe the $(g-2)_\mu$-allowed region.

\section{Discussion and Conclusion}
Both ATLAS and the CMS collaborations searched for new bosons decaying into opposite-sign and different flavor light leptons ($e^\pm\mu^\mp$)~\cite{ATLAS:2019old,CMS:2023pte}.
In the CMS analysis, 
machine-learning techniques are used to enhance the  sensitivity where an excess is observed. ATLAS, on the other hand, did not perform such a dedicated, BDT-optimized resonance search, and did not interpret the results for masses which are different than the SM value of $\sim 125$~GeV.
Therefore, naively, it could be that the CMS analysis is sensitive to a signal hypothesis which was not reachable by ATLAS.
Although a similar excess at 146~GeV is disfavored by ATLAS at $1\sigma$ (as shown in our Fig.~\ref{fig:p2})~\cite{Leney}, it is a ballpark estimate only and not entirely conclusive; a dedicated interpretation of the ATLAS results is required.


Both analyses generated signal samples with two mechanisms: gluon-fusion (ggH) and vector-boson-fusion (VBF).
The contribution of the ggH mechanism to the total cross section is significantly higher~\cite{CMS:2023pte}, and therefore it has the dominant effect on the results.
In order to validate the use of the results by simply comparing cross sections, we compared the kinematic distributions of the leptons between the ggH mechanism and a direct production with leptons from the proton, and found good agreement. 

It is also interesting that CMS reported excesses in the diphoton~\cite{CMS:2023yay} and ditau~\cite{CMS:2022goy} channels at 95 GeV, but only with $2.9\sigma$ local ($1.3\sigma$ global) and $2.6\sigma$ local ($2.3\sigma$ global) significances, respectively. These can be accommodated with an extended Higgs sector~\cite{Biekotter:2023jld, Azevedo:2023zkg, Escribano:2023hxj, Biekotter:2023oen, Bhattacharya:2023lmu}, but a common explanation together with the 146 GeV $e\mu$ excess seems difficult, and  requires further investigation.

In conclusion, the leptophilic 2HDM provides the simplest explanation for the CMS $e\mu$ excess at 146 GeV. It also simultaneously resolves the CDF $W$-mass and the $(g-2)_\mu$ anomalies. A minimal extension of this 2HDM by a singlet charged scalar leads to the Zee model of radiative neutrino mass generation~\cite{Zee:1980ai}. Should the CMS excess be confirmed, a detailed neutrino oscillation fit (similar to what was done in Ref.~\cite{Babu:2019mfe}) with large $Y_{e\mu}$ entry could be performed, which might also lead to concrete predictions in the neutrino sector, including NSI, as well as for charged LFV decays.

\acknowledgments
We thank Saurabh Nangia for valuable technical help with using lepton PDF. AT thanks Julian Heeck for valuable discussion. BD thanks Ashutosh Kotwal and David Toback for convincing him that the CDF $W$-mass anomaly should be taken seriously. BD also thanks the Mitchell Institute at Texas A\&M University for local hospitality, where part of this work was done. BD is supported in part by the US Department of Energy grant No. DE-SC 0017987 and by a URA VSP fellowship. The work of AT is supported in part by the National Science Foundation under Grant PHY-2210428.
YA is supported by the National Science Foundation under Grant No. PHY-2013010. 

\appendix
\section{Muonium-antimuonium oscillation}
The muonium-antimuonium oscillation probability in our 2HDM scenario is given by~\cite{Conlin:2020veq, Fukuyama:2021iyw} 
\begin{equation}
    P(M_\mu\to \overline{M}_\mu) \simeq \frac{64 \alpha^6 m_{\rm red}^6 \tau_\mu^2}{\pi^2} G_{M\overline{M}}^2 \,,
    \label{eq:probmm}
\end{equation}
where $\alpha$ is the fine-structure constant, $m_{\rm red}=m_em_\mu/(m_e+m_\mu)$ is the reduced mass of the electron-muon system, $\tau_\mu$ is the muon lifetime, and $G_{M\overline{M}}$ is the Wilson coefficient which, in our 2HDM scenario, is given by~\cite{Fukuyama:2021iyw} 
\begin{equation}
    G_{M\overline{M}}^2 \simeq 0.32\, \left| \frac{3 G_3}{2} + \frac{G_{45}}{4}  \right|^2 + 0.13\, \left| \frac{G_{45}}{4}-0.68\, G_3  \right|^2 ,
\end{equation}
with the following coefficients in the alignment limit:
\begin{align}
    G_{45} &\equiv - \frac{Y_{e\mu}^{*2} + Y_{\mu e}^2}{8 \sqrt{2}} \left( \frac{1}{m_H^2} - \frac{1}{m_A^2} \right) , \\
    G_{3} &\equiv - \frac{Y_{e\mu}^{*}  Y_{\mu e}} {8 \sqrt{2}} \left( \frac{1}{m_H^2} + \frac{1}{m_A^2} \right)  . 
\end{align} 
We find that for $m_H\simeq m_A$, there is a cancellation in the $G_{45}$ amplitude (at the level of 6\%), while the $G_3$ amplitude vanishes if we consider only $Y_{e\mu}$ (or $Y_{\mu e}$).

\section{Lepton anomalous magnetic moment}
The expression for one-loop contribution of neutral and charged scalars to $(g-2)_\mu$ is given by  
\begin{widetext}
\begin{align}
    \Delta a_\mu & \simeq \frac{m_\mu^2}{16\pi^2}\left[\frac{1}{m_H^2}\left\{\frac{|Y_{e\mu}|^2+|Y_{\mu e}|^2}{6}-2\frac{m_e}{m_\mu}\left(\frac{3}{4}+\log\left(\frac{m_e}{m_H}\right) \right)\Re(Y_{\mu e }Y_{e \mu}) \right\}\right.\nonumber\\
    & \left. \qquad \qquad + \frac{1}{m_A^2}\left\{\frac{|Y_{e\mu}|^2+|Y_{\mu e}|^2}{6}+2\frac{m_e}{m_\mu}\left(\frac{3}{4}+\log\left(\frac{m_e}{m_A}\right) \right)\Re(Y_{\mu e }Y_{e \mu})\right\}
   \right. 
    \nonumber \\
    & \left. \qquad \qquad - \frac{1} {m_{H^+}^2}\frac{|Y_{e\mu}|^2}{6}\right] .
    \label{eq:gm1}
\end{align}
\end{widetext}
In the limit of $m_H\simeq m_A$, the terms proportional to $m_e m_\mu$ cancel. These terms also vanish in the limit of $Y_{\mu e}\to 0$, or if the Yukawa couplings are real. For complex Yukawa couplings, there will be additional strong constraints from electron electric dipole moment~\cite{Roussy:2022cmp}. For our scenario with small $Y_{\mu e}$, Eq.~\eqref{eq:gm1} reduces to the simple expression   
\begin{align}
    \Delta a_\mu  & \simeq \frac{m_\mu^2 |Y_{e\mu}|^2}{96\pi^2 } \left(\frac{1}{m_{H}^2}+\frac{1}{m_{A}^2}-\frac{1}{m_{H^+}^2}\right) .
    \label{eq:gm2}
\end{align}

The same Yukawa coupling $Y_{e\mu}$ also contributes to $(g-2)_e$, and $\Delta a_e$ is given by Eq.~\eqref{eq:gm2} with the replacement $m_\mu\leftrightarrow m_e$. Due to the $m_e^2$ suppression, the corresponding bound on $Y_{e\mu}$ is much weaker. Moreover, it is not clear whether the $(g-2)_e$ result is anomalous. Although the experimental value of $a_e$ has been measured very precisely~\cite{Fan:2022eto}, the SM prediction~\cite{Aoyama:2019ryr} relies on the measurement of the fine-structure constant, and currently there is a $5.5\sigma$ discrepancy between the Paris Rb determination of $\alpha$~\cite{Morel:2020dww} and the Berkeley Cs determination~\cite{Parker:2018vye}. The recent Northwestern result sits in between~\cite{Fan:2022eto}. Until the discrepant $\alpha$ measurements are resolved, we cannot draw any meaningful constraints from $(g-2)_e$.  

\bibliographystyle{utcaps_mod}
\bibliography{bib.bib}

\end{document}